# Enhancing Intrusion Detection In Internet Of Vehicles Through Federated Learning

Abhishek Sebastian
UG Student , B.Tech (ECE)
School of Electronics Engineering
VIT,Chennai.
Chennai , India
Abhishek.sebastian2020@vitstudent.ac.in

Pragna R
UG Student , B.Tech (ECM)
School of Electronics Engineering
VIT,Chennai.
Chennai , India
Pragna.r2020@vitstudent.ac.in

Dr. Sudhakaran G
Assistant Professor
School of Electronics Engineering
VIT,Chennai.
Chennai , India
Sudhakran.g@vit.ac.in

Dr. Renjith P N
Assistant Professor
School of Computer Science and Engineering
VIT,Chennai
Chennai , India
renjith.pn@vit.ac.in

Leela Karthikeyan H
UG Student , B.Tech (ECM)
School of Electronics Engineering
VIT,Chennai
Chennai , India
leelakarthikeyan.h2020@vitstudent.ac.in

*Abstract—* *Federated learning* is a technique of decentralized machine learning that allows multiple parties to collaborate and learn a shared model without sharing their raw data. Our paper proposes a federated learning framework for intrusion detection in Internet of Vehicles (IOVs) using the CIC-IDS 2017 dataset. The proposed framework employs SMOTE for handling class imbalance, outlier detection for identifying and removing abnormal observations, and hyperparameter tuning to optimize the model's performance. We evaluated the proposed framework using various performance metrics and demonstrated its effectiveness in detecting intrusions with other datasets (KDD-Cup 99 and UNSW-NB-15) and conventional classifiers. Furthermore, the proposed framework can protect sensitive data while achieving high intrusion detection performance.

*Keywords— Internet of Vehicles, Intrusion Detection Systems, Federated Learning, Machine Learning.*

## I. INTRODUCTION

With its ability to analyse large amounts of data and identify patterns that humans do not readily recognize; In recent years, machine learning has become increasingly popular as a methodology for intrusion detection. On the other hand, traditional machine learning techniques rely on centralizing data, which poses significant risks to individuals' and organizations' privacy and security. Federated learning is an approach of collaborative machine learning technique in which multiple parties collaborate and learn a shared model without sharing their raw data. This method addresses privacy concerns and allows for the inclusion of data from multiple sources, which can result in improved performance and robustness.

In this paper, we present the following.

1. A federated learning framework for intrusion detection using the CIC-IDS 2017 dataset.

2. The proposed framework employs SMOTE for handling class imbalance, outlier detection for identifying and removing abnormal observations, and hyperparameter tuning to optimize the model's performance.

3. The proposed framework was evaluated using various performance metrics and demonstrated to detect intrusions in different datasets effectively.

## II. LITERATURE REVIEW

Recent studies are looking at the application of federated learning in Internet of Things (IoT) networks for intrusion detection [1-4]. These approaches take advantage of the distributed nature of IoT devices to collect network data and train deep learning models in a decentralized manner. Doing so allows for real-time adaptation to changing network conditions and enhances privacy and security compared to traditional centralized methods. For instance, one approach proposed in [3] uses an efficient communication method and an on-device federated learning technique for deep anomaly detection of time-series data in industrial Internet of Things (IoT), where models are trained on individual IoT devices, reducing communication overhead and preserving privacy. Another ensemble multi-view federated learning technique for intrusion detection in Internet of Things (IoT) networks was presented in [4], which combines multiple views of network data to train an ensemble of federated learning models that detect anomalies in network behavior indicative of security attacks. In [2], a federated learning-based approach to anomaly detection for security attacks in Internet of Things (IoT) networks was proposed, leveraging decentralized data from multiple Internet of Things (IoT) devices to train a federated learning model that detects anomalies in network behavior. Moreover, [5] use deep reinforcement learning to create training data in IOV with prioritized experience and states, as well as a federated learning experience sharing mechanism to protect vehicle privacy.

Other works propose specific applications of federated learning for different tasks, such as resource allocation [6], privacy-preserving learning models for vehicles [7], and outlier detection techniques for machine learning [8-10]. Additionally, the Synthetic Minority Over-Sampling Technique (SMOTE) proposed in [10] addresses the class imbalance in machine learning datasets, while [9] provides a comprehensive survey of outlier detection techniques for temporal data.

Regarding blockchain and federated learning, [14] offers a hybrid architecture that combines a restricted blockchain and a locally directed acyclic graph (DAG) with a deep reinforcement learning-inspired asynchronous federated learning scheme. A blockchain-based federated learning pool (BELP) framework is presented in [15], allowing models to be trained without sharing new data and choosing the most suitable learning. Furthermore, [16] integrates federated learning and local differential privacy (LDP) for crowd-sourcing applications, [17] proposes a federated learning collaborative authentication protocol for shared data to prevent data leakage and reduce the propagation delay of data. Lastly, [18] introduces a semi-synchronous federated learning (Semi-Syn Fed) protocol to improve machine learning performance.

Our article emphasizes the use of the CIC-IDS 2017 dataset [11] for intrusion detection, which incorporates seven publicly accessible shared attack network flows that satisfy real-world requirements. While there is some missing and redundant data in this dataset, which can lead to significant imbalance and poor model performance [12, 13], these difficulties can be overcome by more stringent data preparation.

## III. PROPOSED FRAMEWORK

### A. *Framework Proposed:*

Our study explores the implementation of Federated Learning using two edge devices, namely Internet of Vehicles (IOVs), and a central server. The edge devices train their models locally using their own datasets, and only the trained model is transmitted to the central server. The central server combines the received models and trains a resulting model on its dataset.

To ensure data privacy, the original dataset is partitioned into three distinct parts, with each partition assigned to an edge device and the central server. Each partition adheres to an (80:20) training-testing ratio, which helps protect sensitive data while allowing models to learn from the collective experience of all edge devices. This partitioning principle serves as a secure and efficient framework for learning from distributed datasets, leading to the development of robust models that generalize well across different edge devices.

The models from the edge devices are pickled and transmitted to the central server. The central server trains a super global model with its share of the dataset, and the resulting trained model is then sent back to the edge devices as global model updates. This iterative process continues until satisfactory performance is achieved in the edge devices.

Overall, the study's approach to "Federated Learning" provides a practical solution for addressing the challenge of data privacy in distributed datasets while allowing edge devices (IOV's) to contribute to the collective experience of training robust models.

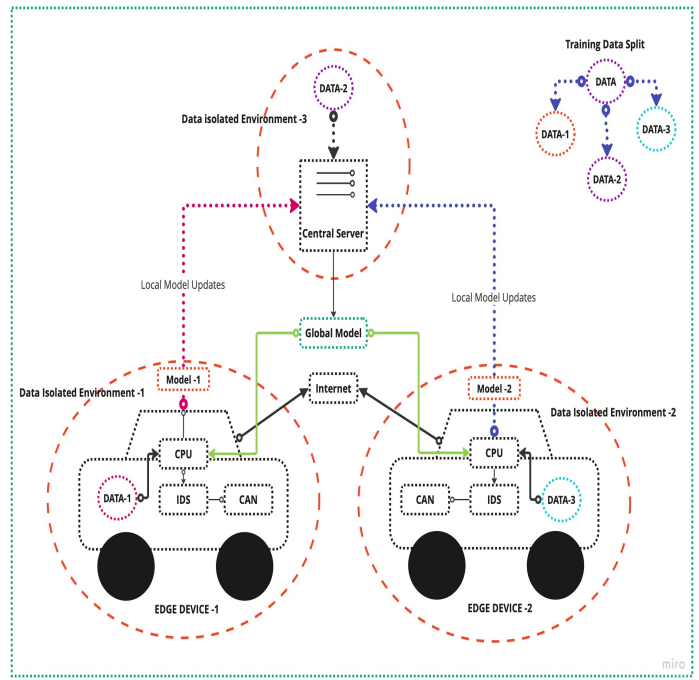

**FIGURE 1** IOV – Federated Learning Architecture

**FIGURE 1.** illustrates the architecture of a federated learning-based Intrusion Detection System (IDS) in an Internet of Vehicles (IoV's) network. The network consists of two Internet-connected vehicles (Edge Devices) that transfer models to a central server. The IDS is placed before the Controller Area Network (CAN) bus to monitor and analyze network traffic. The edge device trains and optimizes the machine learning models using the preprocessed data and sends the model updates to the central server. The central server aggregates the model updates and sends periodic global model updates to the edge devices for retraining. This architecture enables distributed machine learning while maintaining data privacy and enhances the accuracy of IDS in Internet of Vehicles (IoV's) networks.

### B. *Description of the dataset:*

CIC-IDS 2017 (Cyber Intrusion and Cyber Attack Intrusion Detection) is a dataset of network traffic data obtained in a controlled environment that is publicly available. It includes tagged network traffic data that may be used to train and test intrusion detection systems.

For the scope of our research, we used a different variant of the dataset that covers traffic like benign, DoS, port scan, brute force, web attack, bot, and infiltration.

Normal, non-malicious network activity is referred to as benign traffic. On the other hand, DoS attacks seek to make a network resource inaccessible to its intended consumers. Inspecting a computer's open ports for security considerations is referred to as port scanning. To guess passwords or keys, brute force attacks employ trial and error. Web-based systems, applications, or services are the targets of web-based attacks. A *bot* is a computer program that automates chores or performs activities on behalf of a user. Unauthorized access to a network or computer system is referred to as infiltration.

**TABLE 1.** Network Traffic Classes in the dataset "CIC-IDS 2017"

| Si. No | Network Traffic Type | Quantity |
|---|---|---|
| 1 | Benign | 22728 |
| 2 | DoS | 18984 |
| 3 | Port scan | 7946 |
| 4 | Brute force | 2767 |
| 5 | Web attack | 2180 |
| 6 | Bot | 1966 |
| 7 | Infiltration | 36 |

From **TABLE.1**, it can be inferred that the dataset is highly imbalanced (the "Benign" class has the most significant number of samples, while the "Infiltration" class has the smallest number of samples) and can have significant effects on the performance of the machine learning model trained on this dataset. We encounter this issue in the next section with specific pre-processing techniques.

*C. Preprocessing Techniques:*

The proposed framework employs the Synthetic Minority Over-sampling Technique (SMOTE) to address the class imbalance in a dataset composed of seven distinct classes. The dataset consists of 22728 benign instances, 18984 Denial-of-Service (DoS) instances, 7946 Port Scan instances, 2767 Brute Force instances, 2180 Web Attack instances, 1966 Bot instances, and only 36 Infiltration instances. To obtain a balanced data set, we oversampled the minority infiltration class by generating 20000 synthetic instances. Similarly, the other classes, including Port Scan, Brute Force, Web Attack, and Bot, were also oversampled to 20000 instances each. The resulting balanced dataset allows a more accurate and representative evaluation of the classification models. Overall, the SMOTE technique effectively improves the dataset's balance, thereby increasing the reliability of the results obtained from the subsequent analyses.

Despite its effectiveness in balancing imbalanced datasets, oversampling can introduce outliers that may negatively impact the classification models' performance. The proposed framework employs Isolation Trees to identify and remove outliers in the oversampled dataset. Isolation Trees are decision tree that partitions data points based on their isolation depth, which measures the average number of splits required to isolate a data point. The algorithm identifies outliers as instances with high isolation depths, indicating that they are highly distinct from the rest of the data.

**TABLE 2.** Quantity of Network Traffic after Preprocessing

| Si. No | Network Traffic Type | Quantity |
|---|---|---|
| 1 | Benign | 20949 |
| 2 | DoS | 18984 |
| 3 | Port scan | 20000 |
| 4 | Brute force | 20000 |
| 5 | Web attack | 19152 |
| 6 | Bot | 19940 |
| 7 | Infiltration | 9105 |

By removing these outliers, the dataset (**TABLE.2**) is better prepared for further analysis and modelling, as it reduces the risk of overfitting and improves the robustness of the classification models. Overall, combining SMOTE with Isolation Trees provides a reliable and practical approach to handling class imbalance while ensuring the dataset's quality.

## IV. RESULTS AND ANALYSIS

The study utilizes a Cat Boost model, gradient boosting algorithm that utilizes decision trees, as the classifier model for edge devices. The initial hyperparameters for the model are set to a base "depth" of 3, "epochs" of 50, and a "learning rate" of 0.50. To improve the model's accuracy further, the study applies grid search, a hyperparameter tuning technique that exhaustively searches over a specified set of hyperparameters. The search space consists of the "depth," "iterations," and "learning_rate" hyperparameters, where "depth" ranges from 3 to 7, "iterations" ranges from 50 to 200, and "learning_rate" ranges from 0.1 to 1. The resulting model exhibits a significant improvement in accuracy, demonstrating the effectiveness of hyperparameter (**TABLE 3.**) tuning in optimizing deep-learning models for edge devices.

**TABLE 3.** Accuracy of Edge Device Classifiers

| Si. No | Device | Base Model Accuracy | Hyperparameter tuned Model Accuracy |
|---|---|---|---|
| 1 | Edge Device 1 | 94.7 % | 96. 251 % |
| 2 | Edge Device 2 | 95.5 % | 96. 525 % |

In this study, we transfer the best models from edge devices for the central server. It is crucial to note that only the models are transferred, not the data, thereby preserving the fundamental aspect of Federated Learning, i.e., "data privacy." The loaded models from the edge devices are used to create a supermodel utilizing the Bagging Classifier technique. This technique aggregates the predictions of multiple models to produce a single, more accurate model. The resulting supermodel exhibits better robustness than either of the individual edge device models, showcasing the effectiveness of ensemble methods for combining models trained on distributed data. We fit the central server's model using the partitioned dataset previously divided into three parts and validate it accordingly to ensure that the model is accurate and generalizable. The central server aggregates the model updates from the edge devices and broadcasts the new global model, which is then used for further model training on the edge devices. This process is repeated iteratively, allowing the global model to be continuously refined based on the new data collected by the edge devices.

**TABLE 4.** has the network traffic types and their corresponding confusion matrix labels.

**TABLE 4.** Network traffic types - confusion matrix labels

| Network Traffic Type | Confusion Matrix Label |
|---|---|
| Benign | 0 |
| Bot | 1 |
| Brute Force | 2 |
| DoS | 3 |
| Infiltration | 4 |
| Port Scan | 5 |
| Web Attack | 6 |

**TABLE.5**, **TABLE** 6 & **TABLE** 7 are confusion matrices for edge devices (1&2) and central server respectively.

**TABLE 5.** Confusion Matrix for Edge Machine 1

| | | Virtual Edge Machine 1 | | | | | | |
|---|---|---|---|---|---|---|---|---|
| Ground Truth | 0 | 2491 | 90 | 111 | 80 | 7 | 4 | 63 |
| | 1 | 4 | 2616 | 0 | 0 | 0 | 0 | 0 |
| | 2 | 28 | 0 | 2621 | 2 | 0 | 0 | 21 |
| | 3 | 26 | 1 | 5 | 2416 | 2 | 0 | 4 |
| | 4 | 12 | 0 | 0 | 0 | 1197 | 0 | 0 |
| | 5 | 3 | 0 | 0 | 6 | 0 | 2737 | 0 |
| | 6 | 37 | 0 | 134 | 2 | 0 | 1 | 2430 |
| | | 0 | 1 | 2 | 3 | 4 | 5 | 6 |
| | | Prediction | | | | | | |

From the above confusion matrix (**TABLE 5.**), it can be calculated that
  a) Average Precision for edge device 1: 0.9654
  b) Average Recall for edge device 1: 0.9652
  c) Kappa Score for edge device 1: 0.956
  d) Overall Accuracy for edge device 1: 96.229 %

**TABLE 6.** Confusion Matrix for Edge Machine 2

| | | Virtual Edge Machine 2 | | | | | | |
|---|---|---|---|---|---|---|---|---|
| Ground Truth | 0 | 602 | 23 | 32 | 17 | 2 | 0 | 3 |
| | 1 | 0 | 652 | 0 | 0 | 0 | 0 | 0 |
| | 2 | 6 | 0 | 643 | 0 | 0 | 0 | 5 |
| | 3 | 6 | 0 | 3 | 615 | 0 | 0 | 4 |
| | 4 | 1 | 0 | 0 | 0 | 303 | 0 | 0 |
| | 5 | 1 | 0 | 0 | 1 | 0 | 631 | 0 |
| | 6 | 10 | 0 | 28 | 1 | 0 | 0 | 610 |
| | | 0 | 1 | 2 | 3 | 4 | 5 | 6 |
| | | Prediction | | | | | | |

From the above confusion matrix (**TABLE 6.**), it can be calculated that
  a) Average Precision for edge device 2: 0.9689
  b) Average Recall for edge device 2: 0.9689
  c) Kappa Score for edge device 2: 0.96
  d) Overall Accuracy for edge device 2: 96.594 %

**TABLE 7.** Confusion Matrix for Central Server

| | | Central Server | | | | | | |
|---|---|---|---|---|---|---|---|---|
| Ground Truth | 0 | 631 | 26 | 18 | 32 | 4 | 1 | 10 |
| | 1 | 1 | 629 | 0 | 0 | 1 | 0 | 0 |
| | 2 | 16 | 0 | 623 | 1 | 0 | 0 | 2 |
| | 3 | 16 | 0 | 1 | 596 | 0 | 0 | 1 |
| | 4 | 5 | 0 | 0 | 0 | 299 | 0 | 0 |
| | 5 | 2 | 0 | 0 | 2 | 0 | 643 | 0 |
| | 6 | 17 | 0 | 21 | 1 | 0 | 0 | 610 |
| | | 0 | 1 | 2 | 3 | 4 | 5 | 6 |
| | | Prediction | | | | | | |

From the above confusion matrix (**TABLE 7.**), it can be calculated that
  a) Average Precision for central server: 0.963
  b) Average Recall for central server: 0.9624
  c) Kappa Score for central server: 0.953
  d) Overall Accuracy for central server: 95.999 %

The results obtained from the confusion matrices indicate that the proposed federated learning framework demonstrated high accuracy and reliable performance in detecting intrusion attempts. The average precision and recall scores for all three devices were found to be above 0.96, indicating a low rate of false positives and false negatives.

The overall accuracy of the framework was found to be above 95%, suggesting a high level of accuracy in predicting intrusion attempts. The kappa scores obtained were also high, with all devices scoring above 0.95, indicating strong agreement between the predicted and actual classifications. These results suggest that the proposed framework is effective in enhancing the accuracy and reliability of intrusion detection systems in Internet of Vehicles (IoV's) environments.

**TABLE 8.** Accuracy of Different Classifiers in the framework with CIC-IDS 2017 Dataset.

| Sl. No | Device | Accuracy |
|---|---|---|
| 1 | Edge Device 1 | 96.229 % |
| 2 | Edge Device 2 | 96.594 % |
| 3 | Central Sever | 95.999 % |

The above accuracy, precision, recall and confusion matrices are measured concerning testing dataset split, as mentioned earlier in "Proposed Framework."

In the results obtained in **TABLE** 8., the accuracy of the Central Server is slightly less than that of the individual Edge Devices. There are several reasons for this, including the fact that the Central Server model is created by combining the models of the two Edge Devices, which could introduce noise or inconsistencies in the data. Furthermore, we train the Central Server model on a smaller subset of the data than the individual Edge Devices, which could decrease accuracy.

However, Similar Trends (**FIGURE** 2.) are observed in **TABLE** 9. and **TABLE** 10. when the proposed framework is tested on different Datasets (KDD Cup -99 and UNSW-NB-15)

We tested it on other network datasets similar to the CIC-IDS 2017 dataset to further validate the proposed framework and benchmark its performance. Specifically, we conducted experiments on the KDD Cup-99 [20] and UNSW-NB-15 [19] datasets, which also contain network traffic data and share some similarities with the CIC-IDS 2017 dataset. By testing the proposed framework on multiple datasets, we aim to assess its generalizability and effectiveness in a broader context. Through these experiments, we hope to provide a more comprehensive evaluation of the proposed framework and its applicability to a range of network traffic datasets.

Despite different datasets used, similar trends can be observed in the accuracy (**TABLE** 9. & **TABLE** 10.) of the proposed framework when tested on the KDD Cup -99 and UNSW-NB-15 datasets, as shown in **FIGURE** 2. This suggests that the performance of the framework is robust and consistent across multiple datasets, thereby increasing its applicability and reliability in real-world scenarios.

**TABLE 9.** Accuracy of Different Classifiers in the framework with KDD Cup -99 Dataset.

| Si. No | Device | Accuracy |
|---|---|---|
| 1 | Edge Device 1 | 94.98 % |
| 2 | Edge Device 2 | 95.27 % |
| 3 | Central Sever | 94.59 % |

**TABLE 10.** Accuracy of Different Classifiers in the framework with UNSW-NB-15 Dataset.

| Si. No | Device | Accuracy |
|---|---|---|
| 1 | Edge Device 1 | 72.74 % |
| 2 | Edge Device 2 | 71.63 % |
| 3 | Central Sever | 71.53 % |

To ensure consistency in our approach, we followed the same methodology as in the original experiment with the CIC-IDS 2017 dataset. This included using the SMOTE technique for oversampling, Isolation forests for outlier detection, and dividing the data into three sets for each edge device and the central server maintaining data privacy. By following this standardized methodology, we aimed to maintain consistency in the experimental setup and ensure that any observed differences in performance across datasets could be attributed to the nature of the data rather than differences in methodology.

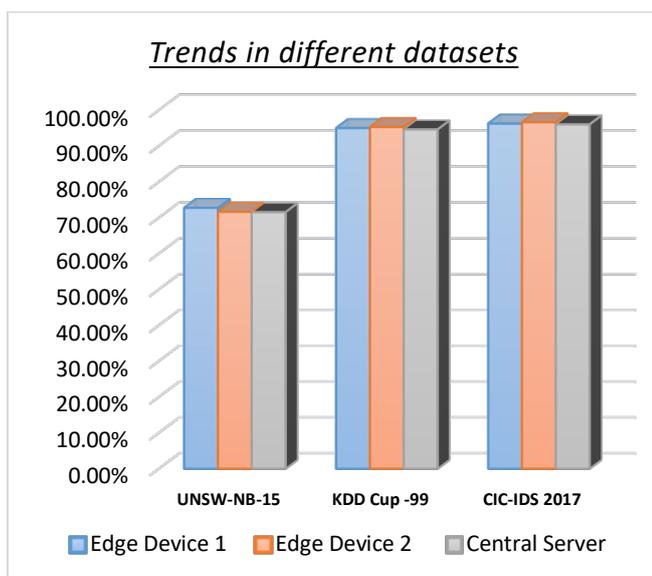

**FIGURE.2** Trends Observed in Different Datasets

**FIGURE.**2 provides a comprehensive view of the comparative performance of the edge devices and the central server in terms of accuracy across the three datasets used in the study, namely CIC-IDS 2017, KDD Cup -99, and UNSW-NB-15.

The proposed framework was developed as an attempt to enhance the existing Intrusion Detection Systems (IDS). The proposed framework was rigorously tested against various conventional classifiers, including Naive Bayes, KNN, Adaboost algorithm, and Gradient Boost algorithm, to assess its efficacy (**TABLE** 11.) in improving the performance of IDS.

**TABLE 11.** Accuracy of Different Classifiers With CIC-IDS 2017 Dataset.

| Si. No | Classifier | Accuracy |
|---|---|---|
| 1 | Proposed Framework | 96.23 % |
| 2 | Random Forest | 94.99% |
| 3 | KNN | 86.21 % |
| 4 | Gradient Boost | 63.68 % |
| 5 | Ada Boost | 59.47 % |
| 6 | Gaussian NB | 40.04% |
| 7 | Multinomial NB | 31.18 % |

**TABLE** 11., presents the accuracy results of different classifiers when tested on the UNSW-NB-15 dataset. The proposed framework achieved the highest accuracy of 96.23%, outperforming all other classifiers by a significant margin. The next best performing classifier was the Random Forest algorithm, with an accuracy of 94.99%. The K-Nearest Neighbor (KNN) algorithm achieved an accuracy of 86.21%, which is significantly lower than that of the proposed framework. The accuracy of the remaining classifiers, including Multinomial NB, Gradient Boost, Ada Boost, and Gaussian NB, was considerably lower, with values ranging from 31.18% to 63.68%.

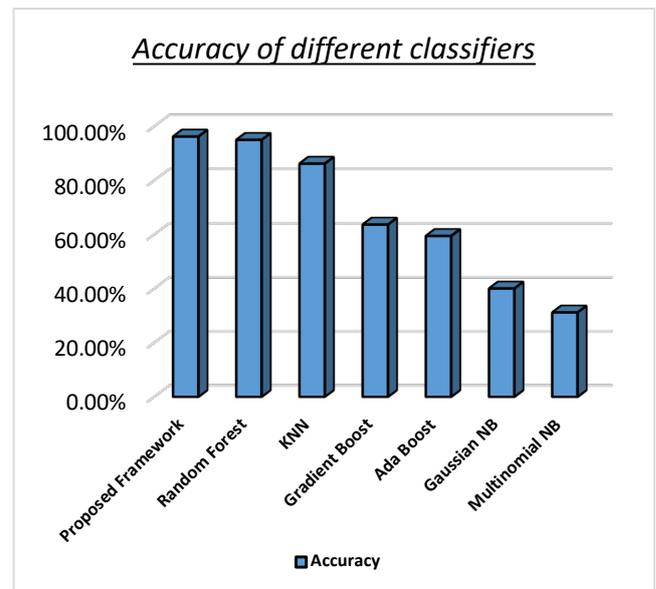

**FIGURE.3** Accuracy of Different Classifiers

These results (**FIGURE.3**) indicate that the proposed framework is a highly effective method for "*enhancing*" the performance of Intrusion Detection Systems. The high accuracy of the proposed framework demonstrates its ability to accurately classify different types of network traffic, which can help in identifying potential security threats and preventing security breaches.

## V. CONCLUSION

In conclusion, this study presents a promising approach for improving the performance of intrusion detection systems in IOVs through federated learning using an imbalanced dataset. The CIC-IDS 2017 dataset, comprising seven distinct classes, was preprocessed using Synthetic Minority Over-sampling Technique (SMOTE) and Isolation Trees to address the class imbalance and to remove outliers. The dataset was partitioned into three parts for two edge devices and one central server to ensure data privacy. The edge devices used a Cat Boost classifier model with hyperparameters optimized through grid search. The resulting model showed a significant increase in accuracy, highlighting the effectiveness of hyperparameter tuning. The edge device models were combined using the Bagging Classifier technique to create a supermodel that demonstrated better robustness. However, the central server model showed slightly less accuracy than the individual edge device models, likely due to noise or inconsistencies in the data or training on a smaller subset. Testing the proposed framework on other network datasets, such as KDD Cup-99 and UNSW-NB-15, provided a comprehensive evaluation of the framework's generalizability. The results showed that it effectively improved accuracy for different network traffic datasets. Furthermore, the proposed framework was compared with existing conventional classifiers, and the results demonstrated that it achieved the highest accuracy among them.Overall, the proposed framework presents a promising approach for federated learning in distributed networks while maintaining data privacy and enhancing the accuracy of intrusion detection systems.

## CODE AVAILABILITY

The code for the proposed framework, which utilizes federated learning for intrusion detection systems on Internet of Vehicles, is available at the GitHub repository: https://github.com/abby1712/Federated_Learning_IDS_On_IOV